\documentstyle[11pt,epsfig]{article}
\date{}
\begin{document}
\title{{\bf Deformed phase space in a two dimensional minisuperspace model}}
\author{H. R. Sepangi$^1$\thanks{email:
 hr-sepangi@sbu.ac.ir}\hspace{1.2mm}, B. Shakerin$^1$ and B. Vakili$^2$\thanks{email: b-vakili@sbu.ac.ir, bvakili45@gmail.com}
\\\\
$^1${\small {\it Department of Physics, Shahid Beheshti University, Evin,
Tehran 19839, Iran }} \\
$^2${\small {\it Department of Physics, Azad University of Chalous, P. O. Box 46615-397, Chalous, Iran}}}
\maketitle
\begin{abstract}
We study the effects of noncommutativity and deformed Heisenberg
algebra on the evolution of a two dimensional minisuperspace
cosmological model in classical and quantum regimes. The phase space
variables turn out to correspond to the scale factor of a flat FRW
model with a positive cosmological constant and a dilatonic field
with which the action of the model is augmented. The exact classical
and quantum solutions in commutative and noncommutative cases are
presented. We also obtain some approximate analytical solutions for
the corresponding classical and quantum cosmology in the presence of
the deformed Heisenberg relations between the phase space variables,
in the limit where the minisuperspace variables are small. These
results are compared with the standard commutative and
noncommutative cases and similarities and differences of these
solutions are discussed. \vspace{5mm}\newline PACS numbers:
98.80.Qc, 04.60.Ds, 04.60.Kz
\end{abstract}

%\baselineskip 24p
\section{Introduction}
As is well known, standard cosmological models based on classical
general relativity have no convincingly precise answer to the
question of the initial conditions from which the universe has
evolved. This can be traced to the fact that these models suffer
from the presence of an initial singularity, the so-called 'Big
Bang' singularity. Indeed, there are various forms of singularity
theorems in general relativity \cite{1}, which show that quite
reasonable assumptions lead to at least one consequence which is
physically unacceptable. Any hope of dealing with such
singularities would be in the development of a concomitant and
conducive quantum theory of gravity \cite{2}. On the other hand,
one of the most important features of theories which deal with
quantum gravity is the existence of a minimal length below which
no other length can be observed \cite{3}. From perturbative string
theory point of view, such a minimal length, of the order of
Planck scale, is due to the fact that strings cannot probe
distances smaller than the string size. Also, the existence of
this minimal length has been suggested in loop quantum gravity
\cite{4}, quantum geometry \cite{5} and black hole physics
\cite{6}. Indeed, at the scale of such a minimum size i.e. the
scales of the order of the Planck length,
$l_p=\sqrt{\frac{G\hbar}{c^3}}$, the quantum effects of
gravitation become as important as the electroweak and strong
interactions. Clearly, at low energy, these quantum gravity
effects are not too important, but in high energy physics, that
is, energies of the order of Planck mass $m_p=\hbar/l_p$ such as
the very early universe or in the strong gravitational fields of a
black hole, one cannot neglect these effects.

One of the most important features of the existence of a minimal
length is that such a length is related to what is
known as the Generalized Uncertainty Principle (GUP); the usual
Heisenberg uncertainty principle should be reformulated at the
Planck scale \cite{7,8}. In a one dimensional system, the simplest
form of the GUP which shows the appearance of a minimum position
uncertainty can be written as \cite{7}
\begin{equation}\label{A}
\bigtriangleup p \bigtriangleup x\geq \frac{\hbar}{2}\left(1+\beta
(\bigtriangleup p)^2+\gamma\right),
\end{equation}
where $\beta$
and $\gamma$ are positive and independent of $\bigtriangleup x$
and $\bigtriangleup p$, but may in general depend on the
expectation values $\langle x\rangle$ and $\langle p\rangle$. If
we take $\gamma=\beta \langle p\rangle^2$, it is possible to
realize equation (\ref{A}) from the following commutation relation
between position and momentum operators
\begin{equation}\label{B} \left[x,p\right]=i\hbar \left(1+\beta
p^2\right).
\end{equation}
In \cite{7}-\cite{9}, more general GUPs are considered. In more than
one dimension GUP naturally implies a noncommutative geometric
generalization of position space \cite{7}. Noncommutativity between
space-time coordinates was first introduced by Snyder \cite{10}
which has lead to a great deal of interest in this area of research
in the recent past \cite{11}.

It is generally an accepted practice to introduce GUP or
noncommutativity either through the coordinates or fields which may
be called geometrical or phase space deformation respectively
\cite{12}-\cite{17}. Applying GUP or noncommutativity to ordinary
quantum field theories where the geometry is considered to obey such
deformations are interesting since they could provide an effective
theory bridging the gap between ordinary quantum field theory and
string theory, currently considered as the most important choice for
quantization of gravity. A different approach to GUP and
noncommutativity is through its introduction in the phase space
constructed by minisuperspace fields and their conjugate momenta
\cite{13}-\cite{17}.  Since cosmology provides the ground for
testing physics at energies which are much higher than those on
Earth, it seems natural to expect the effects of quantum gravity be
observed in this context. Alternatively, in cosmological systems,
since the scale factor, matter fields and their conjugate momenta
play the role of dynamical variables of the system, introducing GUP
and noncommutativity in the corresponding phase space is
particularly relevant.

In general, as we mentioned above, GUP and noncommutativity in
their original form (see \cite{7}) imply a noncommutative
underlying geometry for space-time. However, formulation of
gravity in a non-commutative space-time is highly nonlinear,
rendering the setting up of cosmological models difficult. Here,
our aim is to study the aspects relating to the application of GUP
and noncommutativity in the framework of quantum cosmology, i.e.
in the context of a minisuperspace reduction of dynamics. As is
well-known in the minisuperspace approach to quantum cosmology
which is based on the canonical quantization procedure, one first
freezes a large number of degrees of freedom by the imposition of
symmetries on the spatial metric and then quantizes the remaining
ones. Therefore, in the absence of a full theory of quantum
gravity, quantum cosmology is a quantum mechanical toy model
(finite degrees of freedom) providing a simple arena for testing
ideas and constructions which can be introduced in quantum general
relativity. In this respect, the GUP approach to quantum cosmology
appears to be based on physical grounds. In fact, a generalized
uncertainty principle can be immediately reproduced by deforming
the canonical Heisenberg algebra. In other words, the GUP scheme
relies on a modification of the canonical quantization
prescriptions and, in this respect, can be reliably applied to any
dynamical system. In this sense, one can introduce
noncommutativity between different dynamical variables of the
corresponding minisuperspace and, of course, get different
results. Here, we rely on and use the most common and accepted
practices which have been appearing in the literature over the
past years. It is to be noted that our presentation does not claim
to clear the role of GUP and noncommutativity in cosmology in a
fundamental way since we study the problem in a simple model.
However, this may reflect realistic scenarios in similar
investigations which deal with such problems in a more fundamental
way.

We begin with a flat FRW metric,  a positive cosmological constant
and a homogeneous scalar dilatonic field. We then write the action
in the string frame which leads us to a point like Lagrangian for
the model. We see that the corresponding minisuperspace constructed
by the scale factor $a$ and dilaton field $\phi$ is curvilinear.
Setting up a deformed phase space formalism in such a minisuperspace
is not an easy task. Therefore, we introduce a change of variables
$(a,\phi)\rightarrow (u,v)$ which reduces the minisuperspace to a
linear (Minkowskian) one. These variables are thus suitable
candidates for introducing noncommutativity and GUP in the phase
space of the problem at hand and enable us to present exact
solutions for the classical and quantum commutative and
noncommutative cosmology studied here. Also in the case when the
minisuperspace variables obey the GUP commutation relations, we
obtain approximate analytical solutions for the corresponding
classical and quantum cosmology. Finally, we compare and contrast
these solutions at both classical and quantum levels.
\section{The preliminary setup}
In the pre-big bang scenario, based on the string effective action
\cite{18}, the birth of the universe is described by a transition
from the string perturbative vacuum with weak coupling, low
curvature and cold state to the standard radiation dominated regime,
passing through a high curvature and strong coupling phase. This
transition is made by the kinetic energy term of the dilaton, an
scalar field $\phi$ to which the Einstein-Hilbert action of general
relativity is coupled, see \cite{19} for a more modern review and
\cite{20} for some exact solutions of string dilaton cosmology.
According to this model the lowest order gravi-dilaton effective
action, in the string frame, can be written as \cite{21}
\begin{equation}\label{C}
{\cal S}=-\frac{1}{2 \lambda_s}\int d^4x
\sqrt{-g}e^{-\phi}\left({\cal R}+\partial_{\mu}\phi
\partial^{\mu}\phi-2\Lambda\right),\end{equation}where $\phi (t)$ is
the dilaton field, $\lambda_s$ is the fundamental string length
$l_s$ parameter and $\Lambda$ is a (positive) cosmological constant. We consider a
spatially flat FRW spacetime which is specified by the metric
\begin{equation}\label{D}
ds^2=-dt^2+a^2(t)\delta_{ij}dx^idx^j,
\end{equation}
where $a(t)$ is the scale factor. The Ricci scalar corresponding to metric (\ref{D}) is
\begin{equation}\label{E}
{\cal R}=6\frac{\dot{a}^2}{a^2}+6\frac{\ddot{a}}{a},
\end{equation}
where a dot represents differentiation with respect to $t$.
By substituting (\ref{E}) into (\ref{C}) and integrating over
spatial dimensions, we are led to an effective Lagrangian in
the minisuperspace $Q^A=(a,\phi)$
\begin{equation}\label{F}
{\cal L}=e^{-\phi}\left(6\dot{a}^2a-6\dot{a}a^2\dot{\phi}+a^3\dot{\phi}^2-2\Lambda a^3\right).
\end{equation}
The momenta conjugate to the dynamical variables are given by
\begin{equation}\label{G}
P_a=\frac{\partial {\cal L}}{\partial \dot{a}}=e^{-\phi}(12\dot{a}a-6a^2\dot{\phi}),\hspace{.5cm}
P_{\phi}=\frac{\partial {\cal L}}{\partial \dot{\phi}}=e^{-\phi}(-6\dot{a}a^2+2a^3\dot{\phi}),
\end{equation}
leading to the following Hamiltonian
\begin{equation}\label{H}
{\cal H}=\frac{1}{2}{\cal G}^{AB}P_AP_B+{\cal U}(Q^A)=e^{\phi}\left(-\frac{1}{12a}P_a^2-\frac{1}{2a^3}P_{\phi}^2
+\frac{1}{2a^2}P_aP_{\phi}\right)+2\Lambda a^3e^{-\phi}.
\end{equation}
Now, it is easy to see that the corresponding minisuperspace has
the following minisuper metric
\begin{equation}\label{I}
{\cal G}_{AB}dQ^A dQ^B=e^{-\phi}\left(12a da^2+12a^2 da d \phi +2a^3 d \phi^2\right).
\end{equation}
To apply the deformed commutators to the dynamical variables in a
minisuperspace which is represented by a curved manifold with a
minisuper metric given by (\ref{I}), in a natural generalization,
one can replace $p^2$ in (\ref{B}) with ${\cal G}^{AB}P_AP_B$. In
general, this generalization does not provide a suitable expression
because of the ambiguity in the ordering of factors $Q$ and $P_Q$.
Therefore, the above minisuperspace does not have the desired form
for introducing noncommutativity and GUP among its coordinates. To
avoid the physical difficulties and simplify the model, consider the
following change of variables $Q^A=(a,\phi)\rightarrow q^A=(u,v)$
\begin{equation}\label{J}
u+v=4a^{\alpha}e^{-\phi/2},\hspace{.5cm}u-v=a^{\beta}e^{-\phi/2},
\end{equation}
where $\alpha$ and $\beta$ are two constants which satisfy the relations
\[\alpha+\beta=3,\hspace{.5cm}\alpha \beta=\frac{3}{2}.\]
In terms of these new variables, Lagrangian (\ref{F}) takes the form
\begin{equation}\label{K}
{\cal L}=\dot{u}^2-\dot{v}^2-\omega^2(u^2-v^2),
\end{equation} with the corresponding Hamiltonian becoming
\begin{equation}\label{L}
{\cal H}=\frac{1}{4}\left(p_u^2-p_v^2\right)+\omega^2\left(u^2-v^2\right),
\end{equation}
which describes an isotropic oscillator-ghost-oscillator system
with frequency $\omega^2=\frac{\Lambda}{2}$. Thus, in the
minisuperspace constructed by $q^A=(u,v)$, the metric is
Minkowskian and represented by
\begin{equation}\label{M}
\bar{{\cal G}}_{AB}dq^Adq^B=\frac{1}{2}\left(du^2-dv^2\right).
\end{equation}
Now, we have a set of variables $(u,v)$ endowing the
minisuperspace with a Minkowskian metric and hence
this set of dynamical variables are suitable candidates for
introducing noncommutativity and GUP in the phase space of the
problem at hand. The final remark about the above analysis is that
Lagrangian (\ref{F}) possesses an interesting symmetry thanks to
the presence of the stringy dilaton. This symmetry exhibits itself
through the transformation \cite{21}
\begin{equation}\label{N}
a(t)\rightarrow 1/a(t),\hspace{.5cm}\phi(t)\rightarrow \phi(t)-6\ln a(t).
\end{equation}
It is easy to show that Lagrangian (\ref{F}) is invariant under this
transformation. Such symmetry (duality) is one of the major features
of the solutions of equations of motion in string dilaton cosmology
\cite{20}, so that if the set of variables $(a,\phi)$ solve the
equations of motion, the set $(1/a,\phi-6\ln a)$ is also a solution.
On the other hand, in terms of the variables $(u,v)$ the Lagrangian
takes the simple form (\ref{K}) yielding linear differential
equations for the corresponding dynamical equations. Therefore, the
duality symmetry is nothing but a suitable linear combination of $u$
and $v$. Indeed, one can easily show that Lagrangian (\ref{K}) and
also Hamiltonian (\ref{L}) are invariant under the following
transformations \footnote{In general, Lagrangian (\ref{K}) is
invariant under pseudo rotations in two dimensional Minkowskian
space \[u\rightarrow u\cosh \vartheta +v\sinh
\vartheta,\hspace{.5cm}v\rightarrow u\sinh \vartheta+v\cosh
\vartheta,\]where $\vartheta$ is the parameter of transformations.
In (\ref{O}) and (\ref{P}) we take a special choice for $\vartheta$
to recover the duality of the theory represented by (\ref{N}).}
\begin{equation}\label{O}
u\rightarrow \frac{17}{8}u-\frac{15}{8}v,\hspace{.5cm}v\rightarrow \frac{15}{8}u-\frac{17}{8}v,
\end{equation}
\begin{equation}\label{P}
p_u\rightarrow \frac{17}{8}p_u+\frac{15}{8}p_v,\hspace{.5cm}p_v\rightarrow -
\frac{15}{8}p_u-\frac{17}{8}p_v.
\end{equation}
The preliminary setup for describing the model is now complete. In what
follows we will study the classical and quantum cosmology of the minisuperspace model
described by Hamiltonian (\ref{L}) in noncommutative and GUP frameworks.
\section{Classical model}
As mentioned above, the dynamical system described by Hamiltonian
(\ref{L}) is a simple isotropic oscillator-ghost-oscillator system
and its classical and quantum solutions can be easily obtained.
Since our aim here is to study the effects of deformed Poisson
brackets on the classical trajectories, in what follows we consider
commutative, noncommutative, and GUP classical cosmologies and
compare the results with each other. In the next section we shall
deal with the quantum cosmology of the model.
\subsection{Classical cosmology with ordinary Poisson brackets}
As is well known for a dynamical system with phase space variables
$(q_i,p_i)$, the Poisson algebra is described by the following
Poisson brackets
\begin{equation}\label{Q}
\left\{q_i,q_j\right\}=\left\{p_i,p_j\right\}=0,\hspace{.5cm}\left\{q_i,p_j\right\}=\delta_{ij},\end{equation}where in our case
$q_i(i=1,2)=u,v$ and $p_i(i=1,2)=p_u, p_v$. Therefore, the
equations of motion become
\begin{equation}\label{R}
\dot{u}=\left\{u,{\cal
H}\right\}=\frac{1}{2}p_u,\hspace{.5cm}\dot{p_u}=\left\{p_u,{\cal
H}\right\}=-2\omega^2 u,\end{equation}
\begin{equation}\label{S}
\dot{v}=\left\{v,{\cal
H}\right\}=-\frac{1}{2}p_v,\hspace{.5cm}\dot{p_v}=\left\{p_v,{\cal
H}\right\}=2\omega^2 v.\end{equation}Integrating the above equations, one is led to
\begin{equation}\label{T}
u(t)=\left(u_0^2+\frac{\dot{u}_0^2}{\omega^2}\right)^{1/2}\sin \left[\omega t+\tan^{-1}\frac{\omega u_0}{\dot{u}_0}\right],\end{equation}
\begin{equation}\label{U}
v(t)=\left(v_0^2+\frac{\dot{v}_0^2}{\omega^2}\right)^{1/2}\sin \left[\omega t+\tan^{-1}\frac{\omega v_0}{\dot{v}_0}\right],\end{equation}
where for the initial conditions we take
\begin{equation}\label{V}
\begin{array}{cc}
u(t=0)=u_0, & \dot{u}(t=0)=\dot{u}_0, \\
v(t=0)=v_0,&\dot{v}(t=0)=\dot{v}_0.  \end{array} \end{equation}
The above solutions must satisfy the Hamiltonian constraint,
${\cal H}=0$. Thus, substitution of equations (\ref{T}) and (\ref{U}) into
(\ref{L}) gives the following relation between integration constants
\begin{equation}\label{W}
u_0^2+\frac{\dot{u}_0^2}{\omega^2}=v_0^2+\frac{\dot{v}_0^2}{\omega^2}.
\end{equation}From the above equations, we see that the classical trajectories obey the relation
\begin{equation}\label{X}
v=\pm \cos \left(\tan^{-1}\frac{\omega v_0}{\dot{v}_0}-\tan^{-1}\frac{\omega u_0}{\dot{u}_0}\right)u\pm
\sin \left(\tan^{-1}\frac{\omega v_0}{\dot{v}_0}-\tan^{-1}\frac{\omega u_0}{\dot{u}_0}\right)
\left(u_0^2+\frac{\dot{u}_0^2}{\omega^2}-u^2\right)^{1/2}.
\end{equation}
Note that the minisuperspace of the above model is a
two-dimensional manifold which in terms of the old variables $a$
and $\phi$ is represented by $0<a<\infty$, $-\infty<\phi<+\infty$.
Following \cite{22}, we may divide its boundary into two, the
nonsingular and singular. The nonsingular boundary is the line
$a=0$ with $|\phi|<+\infty$, while at the singular boundary, at
least one of the two variables is infinite. In terms of the
variables $u$ and $v$, introduced in (\ref{J}), the minisuperspace
is recovered by $u>0$, $-u <v<u$, and the nonsingular boundary may
be represented by $u=v=0$. This discussion leads us to the
imposition of more restrictions on the initial conditions
(\ref{V}) such that the classical trajectories would no longer
meet the nonsingular boundary. This condition is achieved when the
coefficient of the second term in (\ref{X}) is nonzero.

Now, let us go back to the old variables $a$ and $\phi$, in terms
of which we obtain the corresponding classical cosmology as
\begin{equation}\label{Y}
4a(t)^{\alpha-\beta}=4a(t)^{\pm \sqrt{3}}=\frac{u+v}{u-v},
\end{equation}
\begin{equation}\label{Z}
\phi(t)=2\beta \ln |a(t)|-2\ln |u-v|,
\end{equation}
leading to the following sets of classical solutions
\begin{equation}\label{AB}
\begin{array}{c}
a_{+}(t)=a_0\left[\tan \omega (t-t_0)\right]^{1/\sqrt{3}}, \\
\\
\phi_{+}(t)=(\sqrt{3}-1)\ln |\tan \omega (t-t_0)|-2\ln |\cos \omega (t-t_0)|+\phi_0,
\end{array}
\end{equation}
and
\begin{equation}\label{AC}
\begin{array}{c}
a_{-}(t)=\left[a_0 \tan \omega (t-t_0)\right]^{-1/\sqrt{3}}, \\
\\
\phi_{-}(t)=(-\sqrt{3}-1)\ln |\tan \omega (t-t_0)|-2\ln |\cos \omega (t-t_0)|+\phi_0,
\end{array}
\end{equation}
where $a_0$, $t_0$ and $\phi_0$ are some constants which can be written in
terms of $u_0$, $\dot{u}_0$, $v_0$ and $\dot{v}_0$. These two sets of
solutions are related to the duality symmetry (\ref{N}), indeed we have
\begin{equation}\label{AD}
a_{-}(t)=\frac{1}{a_{+}(t)},\hspace{.5cm}\phi_{-}(t)=\phi_{+}(t)-6\ln a_{+}(t).
\end{equation}
As we mentioned in the previous section, in the minisuperspace
$(u,v)$ the duality symmetry is denoted by the linear combination
(\ref{O}) of $u$ and $v$. Therefore, applying the duality
transformation (\ref{O}) we are led to the following set of
solutions
\begin{equation}\label{AE}
U(t)=\left(u_0^2+\frac{\dot{u}_0^2}{\omega^2}\right)^{1/2}\left\{\frac{17}{8}\sin
\left[\omega t+\tan^{-1}\frac{\omega u_0}{\dot{u}_0}\right]
-\frac{15}{8}\sin \left[\omega t+\tan^{-1}\frac{\omega v_0}{\dot{v}_0}\right]\right\},\end{equation}
\begin{equation}\label{AF}
V(t)=\left(u_0^2+\frac{\dot{u}_0^2}{\omega^2}\right)^{1/2}\left\{\frac{15}{8}\sin
\left[\omega t+\tan^{-1}\frac{\omega u_0}{\dot{u}_0}\right]
-\frac{17}{8}\sin \left[\omega t+\tan^{-1}\frac{\omega v_0}{\dot{v}_0}\right]\right\}.
\end{equation}
It is clear that these solutions are essentially a special linear
combination of (\ref{T}) and (\ref{U}) which obviously solve the
classical equations of motion because of their linearity.
\subsection{Classical cosmology with noncommutative phase space variables}
Let us now proceed to study the behavior of the above model in a
deformed phase space framework such that the minisuperspace
variables do not (Poisson) commute with each other. In general,
noncommutativity between phase space variables can be understood
by replacing the usual product with the star-product, also known
as the Moyal product law between two arbitrary functions of
position and momentum as
\begin{equation}\label{AG}
(f*_{\alpha}g)(x)=\exp\left[\frac{1}{2}\alpha^{ab}\partial^{(1)}_a
\partial^{(2)}_b\right]f(x_1)g(x_2)|_{x_1=x_2=x},
\end{equation}
where $\alpha^{ab}$ denote the noncommutative parameters \cite{Mo}.
Here, we consider a noncommutative phase space in which the Poisson
algebra is a deformed one given by
\begin{equation}\label{AH}
\left\{q_{inc},q_{jnc}\right\}=\theta \epsilon_{ij},\hspace{.5cm} \left\{p_{inc},p_{jnc}\right\}=0,
\hspace{.5cm}\left\{q_{inc},p_{jnc}\right\}=\delta_{ij},
\end{equation}
where $\epsilon_{ij}$ and $\delta_{ij}$ are Levi-Civita and
Kronecker symbols respectively, $q_{inc}(i=1,2)=u_{nc},v_{nc}$ and
$p_{inc}(i=1,2)=p_{u_{nc}}, p_{v_{nc}}$. With the deformed phase
space defined above, one may consider the Hamiltonian of the
noncommutative model as having the same functional form as
(\ref{L}), but with the dynamical variables satisfying the
above-deformed Poisson brackets, that is
\begin{equation}\label{AI}
{\cal H}_{nc}=\frac{1}{4}\left(p_{u_{nc}}^2-p_{v_{nc}}^2\right)+\omega^2\left(u_{nc}^2-v_{nc}^2\right).\end{equation}
Thus, the dynamics of the system can be described by the following equations of motion
\begin{equation}\label{AJ}
\dot{u_{nc}}=\left\{u_{nc},{\cal
H}_{nc}\right\}=\frac{1}{2}p_{u_{nc}}-2\theta \omega^2v_{nc},\hspace{.5cm}\dot{p_{u_{nc}}}=\left\{p_{u_{nc}},{\cal
H}_{nc}\right\}=-2\omega^2 u_{nc},\end{equation}
\begin{equation}\label{AK}
\dot{v_{nc}}=\left\{v_{nc},{\cal
H}_{nc}\right\}=-\frac{1}{2}p_{v_{nc}}-2\theta \omega^2u_{nc},\hspace{.5cm}\dot{p_{v_{nc}}}=\left\{p_{v_{nc}},{\cal
H}_{nc}\right\}=2\omega^2 v_{nc}.
\end{equation}
Eliminating the momenta from the above equations, we get
\begin{equation}\label{AL}
\ddot{u_{nc}}+\omega^2 u_{nc}+2\theta \omega^2 \dot{v_{nc}}=0,
\end{equation}
\begin{equation}\label{AM}
\ddot{v_{nc}}+\omega^2 v_{nc}+2\theta \omega^2 \dot{u_{nc}}=0.
\end{equation}
We see that the noncommutative parameter appears as a coupling
constant between equations of motion for $u_{nc}$ and $v_{nc}$.
Integrating equations (\ref{AL}) and (\ref{AM}) yields
\begin{equation}\label{AN}
u_{nc}(t)={\cal A} e^{\theta \omega^2 t}\sin\left[\omega \sqrt{1-\theta^2 \omega^2}t+
\delta_1\right]+{\cal B}e^{-\theta \omega^2 t}\sin\left[\omega \sqrt{1-\theta^2 \omega^2}t+\delta_2\right],
\end{equation}
\begin{equation}\label{AO}
v_{nc}(t)=-{\cal A} e^{\theta \omega^2 t}\sin\left[\omega \sqrt{1-\theta^2 \omega^2}t+
\delta_1\right]+{\cal B}e^{-\theta \omega^2 t}\sin\left[\omega \sqrt{1-\theta^2 \omega^2}t+\delta_2\right],
\end{equation}
where ${\cal A}$, ${\cal B}$, $\delta_1$ and $\delta_2$ are integrating constants.
The requirement that the noncommutative Hamiltonian constraints should hold during
the evolution of the system, that is, ${\cal H}_{nc}=0,$ leads to the following relation between integrating constants
\begin{equation}\label{AP}
{\cal A}{\cal B}=0.
\end{equation}
This means that in the noncommutative minisuperspace the system
follows one of the trajectories $u_{nc}=v_{nc}$ (if ${\cal A}=0$)
or $u_{nc}=-v_{nc}$ (if ${\cal B}$=0). In the case when ${\cal
A}=0$, the two coordinates behave similar to two coupled springs,
oscillating  back and forth together like $\rightarrow
\rightarrow$ and $\leftarrow \leftarrow$ with frequency $\omega
\sqrt{1-\theta^2 \omega^2}$ and an exponentially damping
amplitude. Alternatively, if we take ${\cal B}=0$, the two
variables oscillate in opposite directions like $\leftarrow
\rightarrow$ and $\rightarrow \leftarrow$ with the same frequency
$\omega \sqrt{1-\theta^2 \omega^2}$ but with an exponentially
increasing amplitude.

Before going any further some remarks are in order. An important
ingredient in any model theory related to the quantization of a
cosmological setting is the choice of the quantization procedure
used to quantized the system. The most widely used method has
traditionally been the canonical quantization method based on the
Wheeler-DeWitt (WD) equation which is nothing but the application
of the Hamiltonian constraint to the wavefunction of the universe.
A particularly interesting but rarely used approach to study the
quantum effects is to introduce a deformation in the phase space
of the system. It is believed that such a deformation of
phase space is an equivalent path to quantization, in par with
other methods, namely canonical and path integral quantization
\cite{Za}. This method is based on Wigner quasi-distribution
function and Weyl correspondence between quantum mechanical
operators in Hilbert space and ordinary c-number functions in
phase space. The deformation in the usual phase space structure is
introduced by Moyal brackets which are based on the Moyal product
(\ref{AG}) \cite{Mo}. However, to introduce such deformations it
is more convenient to work with Poisson brackets rather than Moyal
brackets.

From a cosmological point of view, models are built in a
minisuperspace. It is therefore safe to say that studying such a
space in the presence of deformations mentioned above can be
interpreted as studying the quantum effects on cosmological
solutions. One should note that in gravity the effects of
quantization are woven into the existence of a fundamental length
\cite{3}. The question then arises as to what form of deformations
in phase space is appropriate for studying quantum effects in a
cosmological model? Studies in noncommutative geometry \cite{11} and
GUP \cite{7} have been a source of inspiration for those who have
been seeking an answer to the above question. More precisely,
introduction of modifications in the structure of geometry in the
way of noncommutativity has become the basis from which similar
modifications in phase space have been inspired. In this approach,
the fields and their conjugate momenta play the role of coordinate
basis in noncommutative geometry \cite{Car}. In doing so an
effective model is constructed whose validity will depend on its
power of prediction.

A question worth asking at this stage is: would the noncommutative
scheme presented above really offer a quantum picture of the model
at hand and should we refrain from using any other quantization
method simultaneously? It is important to note that equivalence
between the two different approaches of quantization cannot hold
true in models where deformation in phase space is introduced in a
Lorentz non-invariant manner, like what we have done here. This is
not hard to understand since the WD equation is a direct
consequence of diffeomorphism invariance and so if a deformation
in phase space breaks such an invariance then the results of
different quantization methods should be different. For models
where the Lorentz invariance deformation is studied see
\cite{Rom}. Therefore, in the next section when we quantize our
model we also invoke noncommutative quantum cosmology based on the
star-product WD equation.
\subsection{Classical cosmology with GUP}
\begin{figure}
\begin{tabular}{ccc} \epsfig{figure=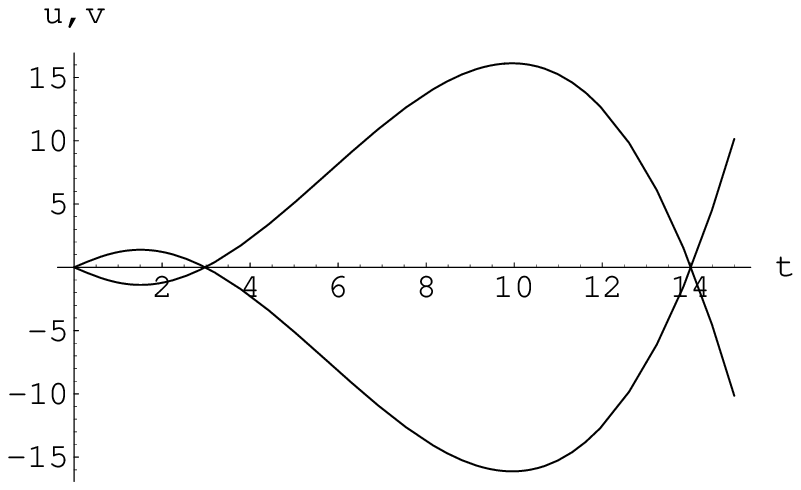,width=7cm}
\hspace{1cm} \epsfig{figure=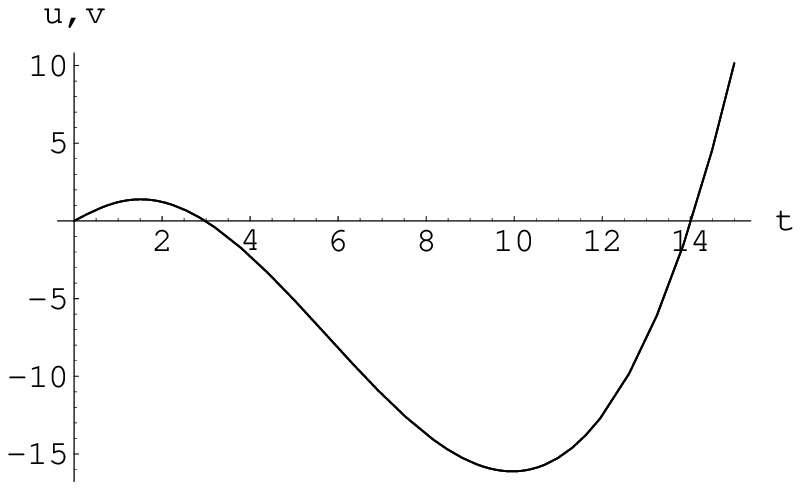,width=7cm}
\end{tabular}
\caption{\footnotesize Approximate behavior of $u(t)$ and $v(t)$ in
the classical GUP framework. The figures are plotted  for numerical
values $\omega=1$ and $\beta=0.01$. We take the initial conditions
$u(t=0)=v(t=0)=0$, $p_u(t=0)=p_v(t=0)=3$ for the figure on the left
and $u(t=0)=v(t=0)=0$, $p_u(t=0)=-p_v(t=0)=3$ for the figure on the
right. } \label{fig1}
\end{figure}

In more than one dimension a natural generalization of equation
(\ref{B}) is defined by the following commutation relations
\cite{7}
\begin{equation}\label{AQ}
\left[x_i,p_j\right]=i\left(\delta_{ij}+\beta
\delta_{ij}p^2+\beta'p_ip_j\right),
\end{equation}
where $p^2=\sum
p_ip_i$ and $\beta,\beta'>0$ are considered as small quantities of
first order. Also, assuming that
\begin{equation}\label{AR}
\left[p_i,p_j\right]=0,\end{equation} the commutation relations
for the coordinates are obtained as
\begin{equation}\label{AS}
\left[x_i,x_j\right]=i\frac{(2\beta-\beta')+(2\beta+\beta')\beta
p^2}{1+\beta p^2}\left(p_ix_j-p_jx_i\right).
\end{equation}
As it is clear from the above expression, the coordinates do not
commute. This means that to construct the Hilbert space
representations one cannot work in the position space. It is
therefore more convenient to work in momentum space. However, since
in quantum cosmology the wavefunction of the universe in momentum
space has no suitable interpretation, we restrict ourselves to the
special case $\beta'=2\beta$. As one can see immediately from
equation (\ref{AS}), the coordinates commute to first order in
$\beta$ and thus a coordinate representation can be defined. Now, it
is easy to show that the following representation of the momentum
operator in position space satisfies relations (\ref{AQ}) and
(\ref{AR}) (with $\beta'=2\beta$) to first order in $\beta$
\begin{equation}\label{AT}
p_i=-i\left(1-\frac{\beta}{3}\frac{\partial^2}{\partial
x_i^2}\right)\frac{\partial}{\partial x_i}.
\end{equation}
Equations (\ref{AQ})-(\ref{AT}) may now be realized from the following
commutation relations between position and momentum operators
\begin{equation}\label{AU}
\left[u,p_u\right]=i\left(1+\beta p^2+2\beta
p_u^2\right),\hspace{.5cm}\left[v,p_v\right]=i\left(1+\beta
p^2+2\beta p_v^2\right),\end{equation}
\begin{equation}\label{AV}
\left[u,p_v\right]=\left[v,p_u\right]=2i\beta p_u
p_v,\end{equation}
\begin{equation}\label{AX}
\left[x_i,x_j\right]=\left[p_i,p_j\right]=0,\hspace{.5cm}x_i(i=1,2)=u,v,\hspace{.5cm}p_i(i=1,2)=p_u,p_v.
\end{equation}
Before quantizing the model within the GUP framework in the next
section, we investigate the effects of the classical version of
GUP, i.e. the classical version of commutation relations
(\ref{AU})-(\ref{AX}) on the above cosmology. As is well known, in
the classical limit the quantum mechanical commutators should be
replaced by the classical Poisson brackets as
$\left[P,Q\right]\rightarrow i \hbar \left\{P,Q\right\}$. Thus, in
classical phase space the GUP changes the Poisson algebra
(\ref{Q}) according to
\begin{equation}\label{AY}
\left\{u,p_u\right\}=1+\beta p^2+2\beta
p_u^2,\hspace{.5cm}\left\{v,p_v\right\}=1+\beta p^2+2\beta
p_v^2,\end{equation}
\begin{equation}\label{AZ}
\left\{u,p_v\right\}=\left\{v,p_u\right\}=2\beta p_u
p_v,\end{equation}
\begin{equation}\label{BA}
\left\{x_i,x_j\right\}=\left\{p_i,p_j\right\}=0,\hspace{.5cm}x_i(i=1,2)=u,v,\hspace{.5cm}p_i(i=1,2)=p_u,p_v,\end{equation}
where $p^2=\frac{1}{2}(p_u^2-p_v^2)$. Such deformed Poisson algebra
is used in \cite{23} to investigate the effects of deformations on
the classical orbits of particles in a central force field and on
the Kepler third law. Also, the stability of planetary circular
orbits in the framework of such deformed Poisson brackets is
considered in \cite{24}. Note that here we deal with modifications
of a classical cosmology that become important only at the Planck
scale where the classical description is no longer appropriate and a
quantum model is required. However, before quantizing the model we
shall provide a deformed classical cosmology. In this classical
description of the universe in transition from commutation relation
(\ref{B}) to its Poisson bracket counterpart we keep the parameter
$\beta$ fixed as $\hbar \rightarrow 0$. In string theory this means
that the string momentum scale is fixed when its length scale
approaches zero. Therefore, the equations of motion read
\begin{equation}\label{BC}
\dot{u}=\left\{u,{\cal H}\right\}=\frac{1}{2}p_u\left(1+5\beta
p^2\right),\hspace{.5cm}\dot{p_u}=\left\{p_u,{\cal
H}\right\}=-2\omega^2u\left(1+\beta p^2+2\beta p_u^2\right)+4\omega^2\beta v p_u p_v,
\end{equation}
\begin{equation}\label{BD}
\dot{v}=\left\{v,{\cal H}\right\}=-\frac{1}{2}p_v\left(1-3\beta
p^2\right),\hspace{.5cm}\dot{p_v}=\left\{p_v,{\cal
H}\right\}=2\omega^2v\left(1+\beta p^2+2\beta
p_v^2\right)-4\omega^2\beta u p_u p_v.
\end{equation}
We see that the deformed classical cosmology forms a system of
nonlinear coupled differential equations which unfortunately cannot
be solved analytically. In figure \ref{fig1}, employing numerical
methods, we have shown the approximate behavior of $u(t)$ and $v(t)$
for typical values of the parameters and initial conditions. As it
is clear from the figure, both variables $u$ and $v$ repeat their
back and forth oscillatory behavior as the noncommutative case, but
here the oscillations are not harmonic. Like the noncommutative
case, depending on the initial conditions, the minisuperspace
variables behave as $\rightarrow \rightarrow$ and $\leftarrow
\leftarrow$, (see the right figure) or $\leftarrow \rightarrow$ and
$\rightarrow \leftarrow$ (see the left figure) i.e. they move back
and forth either in the same or in opposite directions. A comment on
the above results is that although in the limit $\beta \rightarrow
0$ one can recover the ordinary classical cosmology described by
equations (\ref{R}) and (\ref{S}), as this figure shows, taking a
nonzero value for $\beta$ may disturb the oscillatory nature of the
universe. Also, in the presence of $\beta$ terms, the period of
oscillations becomes larger and thus the Big-Crunch in the
corresponding cosmological model occurs later in comparison with the
usual models where $\beta = 0$. This means that the effects of GUP
are important not only in the early but also at late times in the
cosmic evolution. In fact, within the GUP framework, the quantum
gravitational effects may be detected at large scales as well.

\section{The quantum model}
We now focus attention on the study of the quantum cosmology of
the model described above. Here, as in classical cosmology, for
comparison purposes between ordinary commutative, noncommutative,
and GUP frameworks, we consider the quantum cosmology of the model
separately in each case and compare the results. Our starting
point is to construct the WD equation from the corresponding
Hamiltonian.
\subsection{Quantum cosmology with ordinary commutation relations}
\begin{figure}
\begin{tabular}{ccc} \epsfig{figure=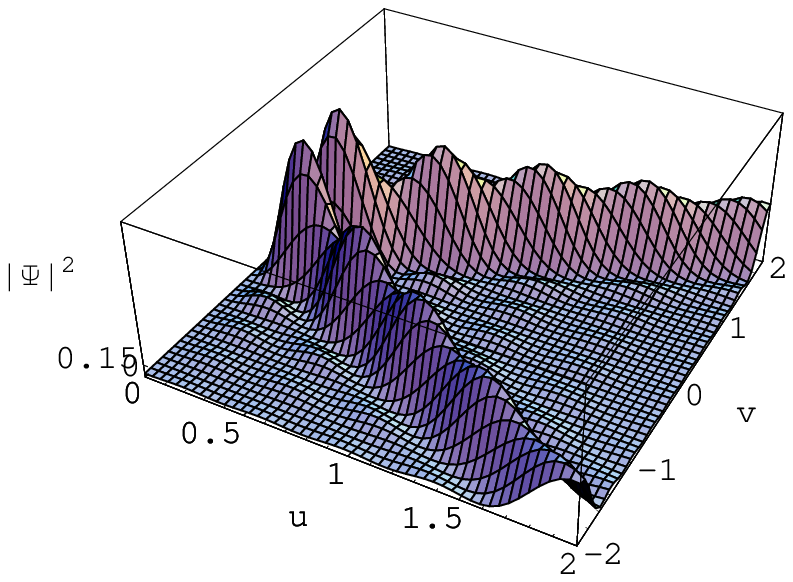,width=9cm}
\hspace{1cm} \epsfig{figure=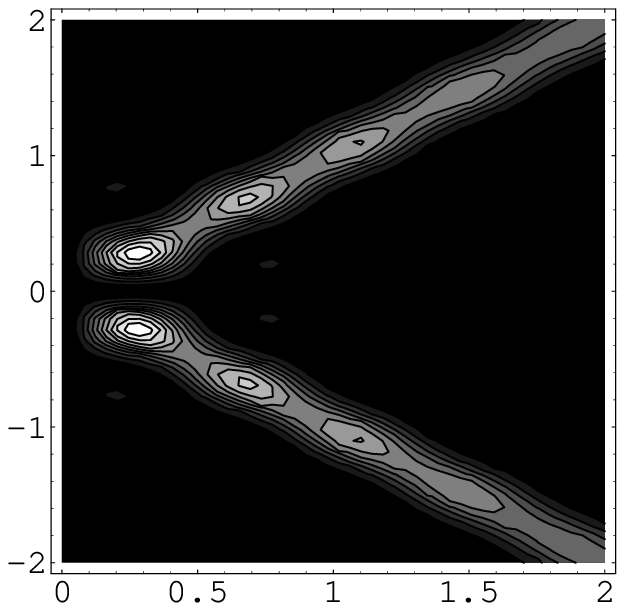,width=6cm}
\end{tabular}
\caption{\footnotesize The figure on the left shows $|\Psi(u,v)|^2$,
the square of the commutative wavefunction while the figure on the
right, the contour plot of $|\Psi(u,v)|^2$. The figures are plotted
for numerical value $\omega=1$ and we have taken a superposition of
eight terms in (\ref{BK}) with all $c_n$ up to $c_{15}$ taken to be
unity.} \label{fig2}
\end{figure}

The quantum version of the model described by relations (\ref{Q})
can be achieved via the canonical quantization procedure which leads
to the WD equation, ${\cal H}\Psi=0$. Here, ${\cal H}$ is the
operator form of the Hamiltonian (\ref{L}) which annihilates the
wavefunction $\Psi$. By replacing  $p_q\rightarrow
-i\frac{\partial}{\partial q}$ in (\ref{L}) we get the WD equation
as
\begin{equation}\label{BE}
\left[-\frac{\partial^2}{\partial u^2}+\frac{\partial^2}{\partial v^2}+
4\omega^2(u^2-v^2)\right]\Psi(u,v)=0.
\end{equation}
This equation is a quantum isotropic oscillator-ghost-oscillator system
with zero energy. Therefore, its solutions belong to a subspace of the
Hilbert space spanned by separable eigenfunctions of a two dimensional
isotropic simple harmonic oscillator Hamiltonian. Separating the
eigenfunctions of (\ref{BE}) in the form
\begin{equation}\label{BF}
\Psi_{n_1,n_2}(u,v)=U_{n_1}(u)V_{n_2}(v),\end{equation}
yields
\begin{equation}\label{BG}
U_{n_1}(u)=\left(\frac{2\omega}{\pi}\right)^{1/4}\frac{e^{-\omega u^2}}{\sqrt{2^{n_1}n_1!}}H_{n_1}(\sqrt{2\omega}u),
\end{equation}
\begin{equation}\label{BH}
V_{n_2}(v)=\left(\frac{2\omega}{\pi}\right)^{1/4}\frac{e^{-\omega v^2}}{\sqrt{2^{n_2}n_2!}}H_{n_2}(\sqrt{2\omega}v),
\end{equation}
subject to the restriction $n_1=n_2=n$. In (\ref{BG}) and
(\ref{BH}), $H_n(x)$ are Hermite polynomials and the
eigenfunctions are normalized according to
\begin{equation}\label{BI}
\int_{-\infty}^{+\infty}e^{-x^2}H_n(x)H_m(x)dx=2^n \pi^{1/2}n! \delta_{mn}.
\end{equation}
Now, we impose the boundary condition on these solutions such that
at the nonsingular boundary (at $u=v=0$) the wavefunction vanishes
\cite{25}, that is, $\Psi(u=0,v=0)=0$, which yields
\begin{equation}\label{BJ}
H_n(u=0)H_n(v=0)=0\Rightarrow n=\mbox{odd}.
\end{equation}
In general, one of the most important features in quantum cosmology
is the recovery of classical cosmology from the corresponding
quantum model, or in other words, how can the WD wavefunctions
predict a classical universe. In this approach, one usually
constructs a coherent wavepacket with good asymptotic behavior in
the minisuperspace, peaking in the vicinity of the classical
trajectory.  Therefore, we may now write the general solution of the
WD equation as a superposition of the above eigenfunctions
\begin{equation}\label{BK}
\Psi(u,v)=\left(\frac{2\omega}{\pi}\right)^{1/2}e^{-\omega(u^2+v^2)}\sum_{n=odd}
\frac{c_n}{2^nn!}H_n(\sqrt{2\omega}u)H_n(\sqrt{2\omega}v).
\end{equation}
Figure \ref{fig2} shows the square of the wavefunction and its
contour plot. As we can see from this figure the peaks follow a
path which can be interpreted as the classical trajectories
(\ref{X}). The crests are symmetrically distributed around $v=0$
which correspond to the $\pm$ signs in (\ref{X}). Thus, it is seen
that there is an almost good correlations between the quantum
patterns and classical trajectories in the $u-v$ plane.

\subsection{Noncommutative quantum model}
Now, the quantum version of the noncommutative cosmology is
achieved by replacing the Poisson brackets with the corresponding
Dirac commutators, $\left\{,\right\}\rightarrow -i\left[,\right]$.
Thus, the Poisson brackets (\ref{AH}) between minisuperspace
variables should be modified as follows
\begin{equation}\label{BL}
\left[u_{nc},v_{nc}\right]=i\theta,\hspace{.5cm}\left[u_{nc},p_u\right]=\left[v_{nc},p_v\right]=i.
\end{equation}
The corresponding quantum cosmology can be obtained by modification
of the operator product in the WD equation ${\cal H}\Psi=0$ with the
Moyal deformed product ${\cal H}_{nc}*\Psi=0$, where ${\cal H}_{nc}$
is the noncommutative Hamiltonian (\ref{AI}). However, there is an
alternative expression for the Moyal star product which is given by
the shift formula
\begin{equation}\label{BM}
u_{nc}=u-\frac{i}{2}\theta p_v,\hspace{.5cm}v_{nc}=v+\frac{i}{2}\theta p_u,
\end{equation}
\begin{equation}\label{BN}
p_{u_{nc}}=p_u,\hspace{.5cm}p_{v_{nc}}=p_v.
\end{equation}
It can easily be checked that if the noncommutative variables obey
relations (\ref{BL}), then $(u,v,p_u,p_v)$ satisfy the usual
Heisenberg algebra
\begin{equation}\label{BO}
\left[u,v\right]=0,\hspace{.5cm}\left[u,p_u\right]=
\left[v,p_v\right]=i,\hspace{.5cm} \left[u,p_v\right]=\left[v,p_u\right]=0.
\end{equation}
In terms of these commutative variables the Moyal WD equation ${\cal
H}_{nc}*\Psi=0$ transforms to the usual WD equation ${\cal H}\Psi=0$
with Hamiltonian
\begin{equation}\label{BP}
{\cal H}=\frac{1}{4}\left(1-\omega^2 \theta^2\right)\left(p_u^2-p_v^2\right)+
\omega^2\left(u^2-v^2\right)-\theta \omega^2\left(up_v+vp_u\right).
\end{equation}
Therefore, the noncommutative version of the WD
equation can be written as
\begin{equation}\label{BR}
\frac{1}{1-\omega^2\theta^2}{\cal H}\Psi(u,v)=H\Psi(u,v)=
\left[\frac{1}{4}\left(p_u^2-p_v^2\right)+\Omega^2\left(u^2-v^2\right)-
\theta \Omega^2\left(up_v+vp_u\right)\right]\Psi(u,v)=0,
\end{equation}
where
\begin{equation}\label{BS}
\Omega^2=\frac{\omega^2}{1-\theta^2\omega^2}.
\end{equation}
The Hamiltonian operator consists of two parts
\begin{equation}\label{BT}
H=H_{uv}-\theta \Omega^2L_{uv},
\end{equation}
where
\begin{equation}\label{BU}
H_{uv}=\frac{1}{4}\left(p_u^2-p_v^2\right)+\Omega^2\left(u^2-v^2\right),\hspace{.5cm}
\mbox{and}\hspace{.5cm}L_{uv}=up_v+vp_u.
\end{equation}
It is easy to see that these two operators commute with each other
\begin{equation}\label{BV}
\left[H_{uv},L_{uv}\right]=0,
\end{equation}
which means that $H_{uv}$ and $L_{uv}$ have simultaneous eigenfunctions
\footnote{Like the theory of ordinary harmonic oscillator, we can also
define the creation and annihilation operators
\[\hat{a}_u=\sqrt{\Omega}u+\frac{i}{2\sqrt{\Omega}}p_u,\hspace{.5cm}
\hat{a}^{\dag}_u=\sqrt{\Omega}u-\frac{i}{2\sqrt{\Omega}}p_u,\] and
similarly for $\hat{a}_v$ and $\hat{a}^{\dag}_v$, satisfying the
following commutation relations
\[\left[\hat{a}_u,\hat{a}^{\dag}_u\right]=\left[\hat{a}_v,\hat{a}^{\dag}_v\right]=1,\] with
other commutators being zero. In terms of these operators $H_{uv}$
and $L_{uv}$ can be viewed as \[H_{uv}=\Omega
\left(\hat{a}^{\dag}_u
\hat{a}_u-\hat{a}^{\dag}_v\hat{a}_v\right),\hspace{.5cm}L_{uv}=i\left(\hat{a}^{\dag}_u
\hat{a}^{\dag}_v-\hat{a}_u\hat{a}_v\right).\] Use of the same
commutators as above between $\hat{a}$ and $\hat{a}^{\dag}$ makes
it easy to see that $\left[H_{uv},L_{uv}\right]=0$.}. In the
coordinate system $(u,v)$ the WD equation (\ref{BR}) is not
separable. Thus, to transform it to a separable equation, consider
the new variables $(\rho,\varphi)$ defined as
\begin{equation}\label{BX}
u=\rho \cosh \varphi,\hspace{.5cm}v=\rho \sinh \varphi,
\end{equation}
in terms of which we have
\begin{equation}\label{BY}
L_{uv}=-iu\frac{\partial}{\partial v}-iv\frac{\partial}{\partial u}=-i\frac{\partial}{\partial \varphi},
\end{equation}
\begin{equation}\label{BZ}
H_{uv}=-\frac{1}{4}\left(\frac{\partial^2}{\partial u^2}-\frac{\partial^2}{\partial v^2}\right)+
\Omega^2\left(u^2-v^2\right)=
-\frac{1}{4}\left(\frac{\partial^2}{\partial \rho^2}+\frac{1}{\rho}\frac{\partial}{\partial \rho}-
\frac{1}{\rho^2}\frac{\partial^2}{\partial \varphi^2}\right)+\Omega^2\rho^2.
\end{equation}
Therefore, if we separate the solution of the WD equation according
to $\Psi(\rho,\varphi)=R(\rho)e^{i\nu \varphi}$, it is easy to see
that this is an eigenfunction of $L_{uv}$ with eigenvalue $\nu$. The
requirement that $\Psi(\rho,\varphi)$ should solve the WD equation
yields the following differential equation for $R(\rho)$
\begin{equation}\label{CA}
\frac{d^2R}{d\rho^2}+\frac{1}{\rho}\frac{dR}{d\rho}+\left(4\nu \Omega^2 \theta-4\Omega^2 \rho^2+\frac{\nu^2}{\rho^2}\right)R=0.
\end{equation}
This equation, after a change of variable $r=2\Omega \rho^2$ and transformation $R=\frac{1}{\rho}{\cal W}$, becomes
\begin{equation}\label{CB}
\frac{d^2{\cal W}}{dr^2}+\left(-\frac{1}{4}+\frac{\kappa}{r}+\frac{1/4-\mu^2}{r^2}\right){\cal W}=0,
\end{equation}
where $\kappa=\frac{1}{2}\nu \Omega \theta$ and $\mu=i\frac{\nu}{2}$. The above equation is the Whittaker differential
equation and its solutions can be written in terms of confluent hypergeometric functions
$M(a, b; x)$ and $U(a, b; x)$ as
\begin{equation}\label{CD}
{\cal W}(r)=e^{-r/2}r^{\mu+\frac{1}{2}}\left[cU(\mu-\kappa+\frac{1}{2},2\mu+1;r)+c'M(\mu-\kappa+
\frac{1}{2},2\mu+1;r)\right].
\end{equation}
In view of the asymptotic behavior of $M(a, b; x) \sim e^x/x^{b-a}$
\cite{26}, we take $c' = 0$. Thus, going back, the eigenfunctions of
the WD equation can be written in terms of variables
$(\rho,\varphi)$ as
\begin{equation}\label{CE}
\Psi_{\nu}(\rho,\varphi)=\rho^{i\nu}e^{-\Omega \rho^2}U\left(\frac{1-\nu \theta \Omega}{2}+
i\frac{\nu}{2},1+i\nu ;2\Omega \rho^2\right)e^{i\nu \varphi}.
\end{equation}
If we demand that $\Psi_{\nu}$ should be a single-valued function of
$\varphi$, then $\nu$ should be an integer $\nu=n$. Now, we may
write the noncommutative wavefunction in the $(\rho,\varphi)$
coordinates as
\begin{equation}\label{CF}
\Psi(\rho,\varphi)=\sum_{n=-\infty}^{+\infty}C_n\rho^{in}e^{-\Omega \rho^2}U
\left(\frac{1-n \theta \Omega}{2}+i\frac{n}{2},1+in ;2\Omega \rho^2\right)e^{in \varphi}.
\end{equation}
Note that in terms of variables  $(\rho,\varphi)$, the boundary
condition $\Psi(u=0,v=0)=0$ takes the form $\Psi(\rho=0,\varphi)=0$
which for (\ref{CF}) is automatically held.

\begin{figure}
\begin{tabular}{ccc} \epsfig{figure=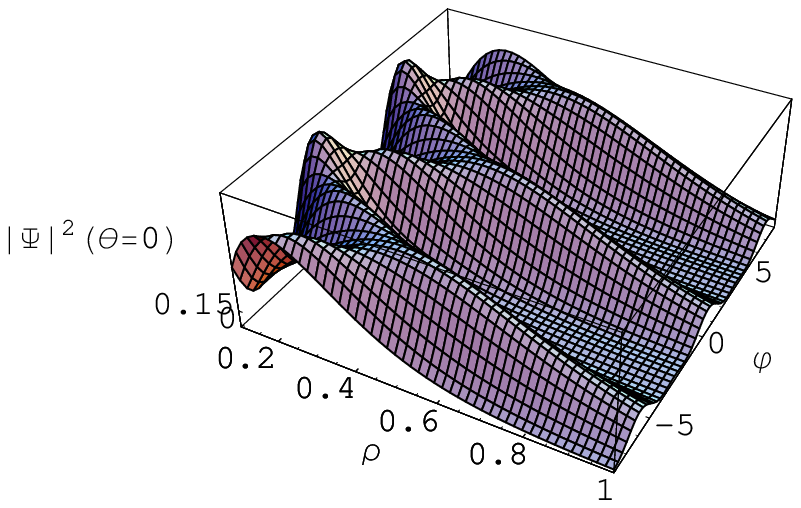,width=9cm}
\hspace{1cm} \epsfig{figure=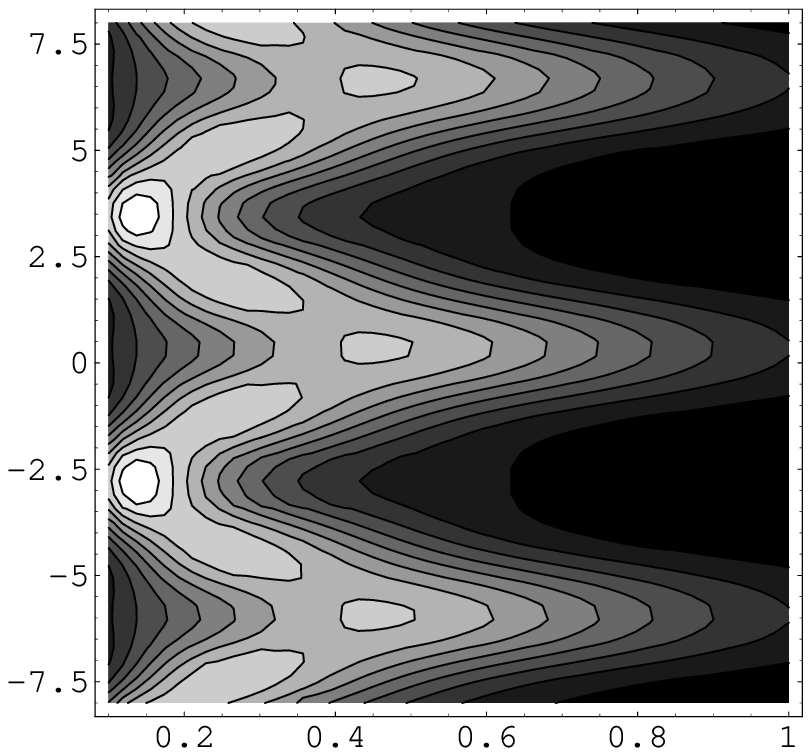,width=6cm}
\end{tabular}
\begin{tabular}{ccc} \epsfig{figure=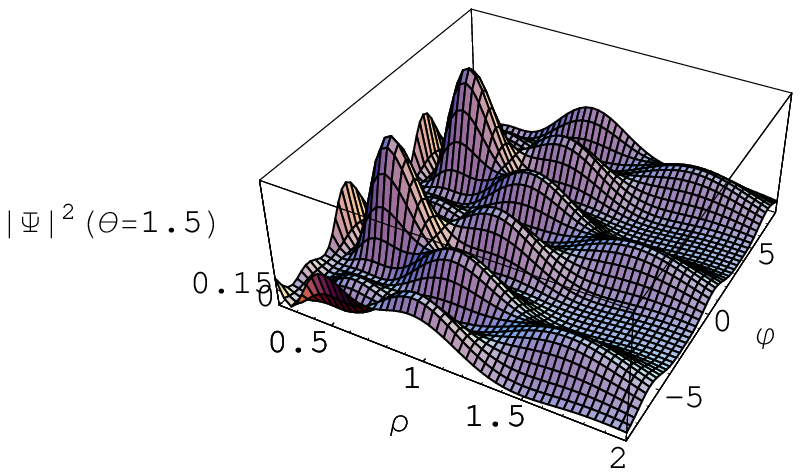,width=9cm}
\hspace{1cm} \epsfig{figure=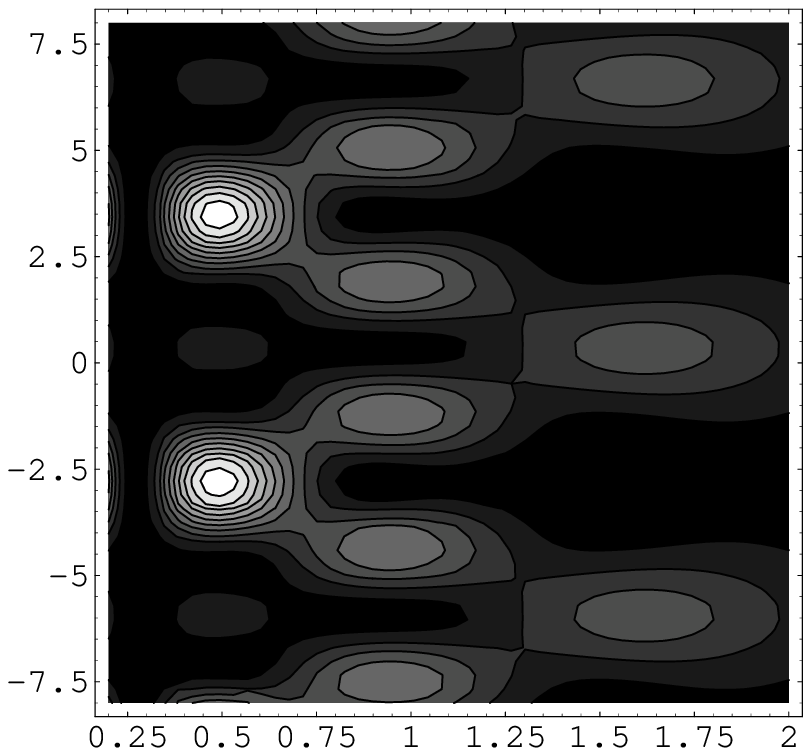,width=6cm}
\end{tabular}
\caption{\footnotesize Up: the figure on the left shows
$|\Psi(u,v)|^2$, the square of the commutative wavefunction in
$(\rho,\varphi)$ coordinates while the figure on the right shows its
contour plot. Bottom: the same figures in the noncommutative case.
The figures are plotted for the numerical value $\Omega=1$ and we
have taken a superposition of three terms in (\ref{CF}) with
$C_1=C_2=C_3=1$.} \label{fig3}
\end{figure}
In figure \ref{fig3} we have plotted the wavefunction (\ref{CF})
in the noncommutative quantum model. In the case where $\theta=0$
(commutative model), although we have analyzed the behavior of the
wavefunction in the previous subsection, we have also considered
it again in the $(\rho,\varphi)$ coordinates. As we can see from
this figure, the commutative wavefunction has two dominant peaks
in the vicinity of $\rho=0$ which then follow a path that can be
interpreted as the classical trajectory (compare with the similar
behavior in figure \ref{fig2}). Therefore, like the commutative
wavefunction in the $(u,v)$ coordinates, there is also a
correlation between the classical and quantum schemes. On the
other hand, the noncommutative wavefunction predicts the emergence
of the universe from a state corresponding to one of the two
dominant peaks. Although there are some small peaks in this
figure, as $\rho$ grows, their amplitude are suppressed. We see
that the correlation with classical trajectories is missed, i.e.
the noncommutativity implies that the universe escapes the
classical trajectories and approaches a stationary state.
\subsection{Quantum model with GUP}
The study of quantum cosmology of the model presented above in the
GUP framework is the goal we shall pursue in this subsection. The
Hamiltonian of the model is given by (\ref{L}) and the corresponding
commutation relations between dynamical variables by equations
(\ref{AU})-(\ref{AX}). In the WD equation we take the representation
(\ref{AT}) for the momenta $p_u$ and $p_v$ which leads to the
following equation up to first order in $\beta$
\begin{equation}\label{CG}
\left[\frac{2}{3}\beta \frac{\partial^4}{\partial u^4}-\frac{\partial^2}{\partial u^2}-
\frac{2}{3}\beta \frac{\partial^4}{\partial v^4}+\frac{\partial^2}{\partial v^2}+
4\omega^2(u^2-v^2)\right]\Psi(u,v)=0.
\end{equation}
We again separate the solutions in the form $\Psi(u,v)=U(u)V(v)$,
leading to
\begin{equation}\label{CH}
-\frac{2}{3}\beta \frac{d^4W_i}{dw_i^4}+\frac{d^2W_i}{dw_i^2}+(\lambda-4\omega^2 w_i^2)W_i=0,
\end{equation}
where $W_i(i=1,2)=U,V$, $w_i(i=1,2)=u,v$ and $\lambda$ is the
separation constant. The appearance of a fourth order differential
equation to describe a physical phenomena is interesting, since it
requires investigation of the corresponding boundary conditions
which is not the goal of our study in this paper. These equations
cannot be solved analytically, but since $\beta$ is a small
parameter which appears only in the fourth order term, we may look
for an approximation method which, within its domain of validity,
leads us to a second order differential equation. From figure
\ref{fig2}, we note that  the dominant peaks of the wavefunction
occur in the vicinity of $u,v\sim 0$. On the other hand, the
effects of $\beta$ are important at the Planck scales, or in
cosmology language in the very early times of the cosmic
evolution, which in our model means $u,v \sim 0$. When $\beta=0$,
the solutions of equation (\ref{CH}) for $\lambda/{2\omega}=2n+1$
are given by (\ref{BG}) and (\ref{BH}). In the limit
$u,v\rightarrow 0$ we may take $e^{-\omega (u^2+v^2)}\sim 1$ which
means that $W_i\propto H_n(\sqrt{2\omega}w_i)$. Therefore, in this
limit, we have
\begin{equation}\label{CI}
\frac{d^4W_i}{dw_i^4}=\left(\frac{2\omega}{\pi}\right)^{1/4}
\frac{64\omega^2 n(n-1)(n-2)(n-3)}{\sqrt{2^n n!}}H_{n-4}(\sqrt{2\omega}w_i),
\end{equation}
where we have used the relation $dH_n(x)/dx=2nH_{n-1}(x)$. Now,
using the following series formula for Hermite polynomials
\begin{equation}\label{CJ}
H_n(x)=\sum_{s=0}^{[n/2]}(-1)^s\frac{n!}{(n-2s)!s!}(2x)^{n-2s},
\end{equation}
we are led to the approximate formula $H_{n-4}(x)\propto H_n(x)$
in the case of a small argument and $n\geq 4$. Thus, in our
approximation, equation (\ref{CH}) becomes
\begin{equation}\label{CK}
\frac{d^2W_i}{dw_i^2}+2\omega\left(2n+1-2\beta_0-2\omega w_i^2\right)W_i=0,
\end{equation}
where $\beta_0$, up to a numerical factor, is
\[\beta_0=\frac{32}{3}\beta \omega n(n-1)(n-2)(n-3).\] The
solutions of the above equation up to a normalization factor can
be written as
\begin{equation}\label{CL}
W_n(w)=e^{-\omega w^2}H_{n-\beta_0}(\sqrt{2\omega}w).
\end{equation}
Finally, the general solutions of the WD equation (\ref{CG}) for
small values of $u$ and $v$ are as follows
\begin{equation}\label{CM}
\Psi(u,v)=\sum_{n=odd}C_n e^{-\omega(u^2+v^2)}H_{n-\beta_0}(\sqrt{2\omega}u)H_{n-\beta_0}(\sqrt{2\omega}v),
\end{equation}
where to recover solution (\ref{BK}) in the limit
$\beta_0\rightarrow 0$, the summation is taken over the odd values
of $n$. Also, to satisfy the boundary condition $\Psi(u=0,v=0)=0$,
we choose $\beta_0$ such that $n-\beta_0$ is an odd integer for
$n>4$.

\begin{figure}
\begin{tabular}{ccc} \epsfig{figure=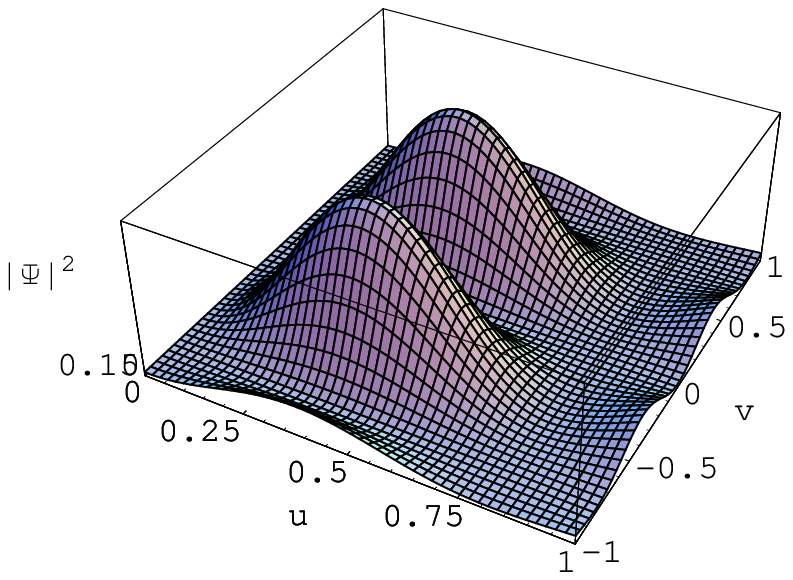,width=9cm}
\hspace{1cm} \epsfig{figure=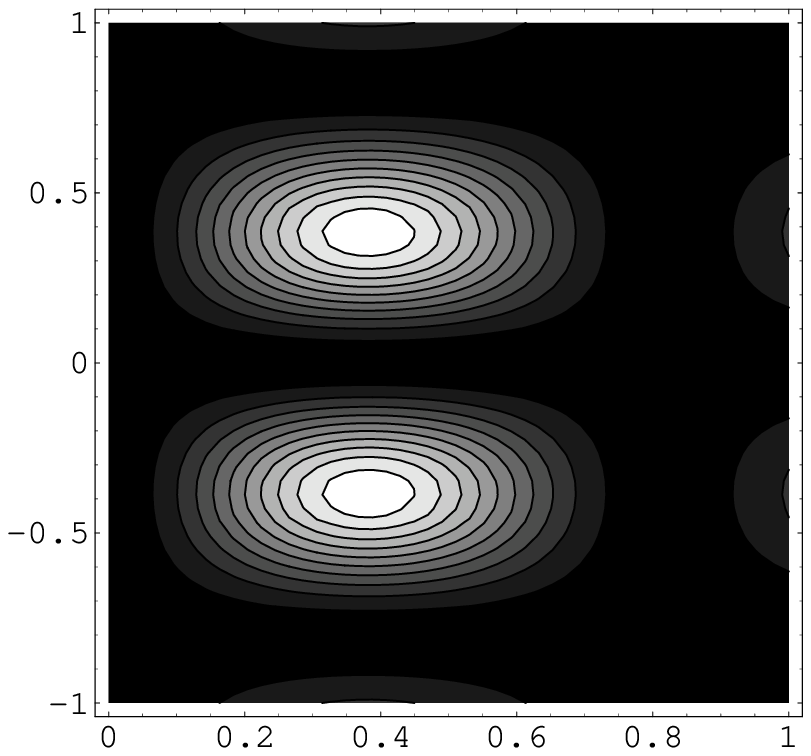,width=6cm}
\end{tabular}
\caption{\footnotesize The figure on the left shows $|\Psi(u,v)|^2$,
the square of the GUP wavefunction while the figure on the right,
the contour plot of $|\Psi(u,v)|^2$. The figures are plotted for
numerical value $\omega=1$, $\beta_0=4$ and we have taken a
superposition of four terms in (\ref{CM}) with all $C_n$ taken to be
unity.} \label{fig4}
\end{figure}
Figure \ref{fig4} shows the wavefunction of the corresponding
universe when the minisuperspace variables obey the GUP relations
for small values of $u$ and $v$. As is clear from this figure the
wavefunction has two single peaks which are symmetrically
distributed around $v=0$. Compare to the commutative wavefunction
(see figure \ref{fig2}), here we have no wave packet with peaks
following the classical trajectories. We see that instead of a
series of peaks in the ordinary WD approach, we have only a couple
of dominant peaks. This means that, similar to the noncommutative
case and within the context of the GUP framework, the wavefunction
also shows a stationary behavior. One may then conclude that from
the point of view adopted here, noncommutativity and GUP may have
close relations with each other.

We should note that the above analysis on the behavior of the GUP
wavefunction is achieved in the region $u,v\sim 0$. If we relax
this approximation, equation (\ref{CI}) is no longer valid and in
the perturbation method we should use  solutions (\ref{BG}) and
(\ref{BH}) in the $\beta$-term of equation (\ref{CH}) in order to
obtain the GUP wavefunction. However, since for large values of
$u$ and $v$ these solutions have an exponentially decreasing
behavior, we do not expect a major effect on the behavior of the
above wavefunction even if we consider the problem more
rigorously. We can also rely on the above approximate GUP
solutions for a wider range of $u$ and $v$.

\begin{figure}\begin{center}
\epsfig{figure=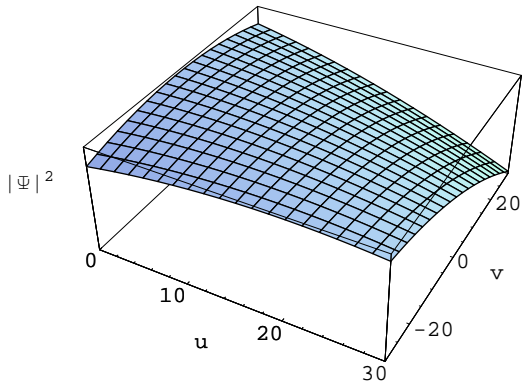,width=7cm} \caption{\footnotesize The figure
shows $|\Psi(u,v)|^2$, the square of the GUP wavefunction
(\ref{CY}). The figure is plotted for numerical values $\omega=1/24$,
$\beta=8$ and we have taken a superposition of four terms in
(\ref{CY}) with all $C(E)$ taken to be unity.}
\label{fig5}\end{center}
\end{figure}

To clarify this point we investigate the quantum GUP model in a
different representation. To do this, we remember that the WD
equation (\ref{CG}) is based on the representation (\ref{AT}) of
momenta in the $u-v$ space. In this sense we note that the
existence of a minimal length means that we cannot have localized
physical states. However, as is shown in \cite{7}, when there is
no minimal uncertainty in momentum one can work with the momentum
space wavefunction $\Phi(\vec{p})$ with the following
representation \cite{7}
\begin{equation}\label{CN}
p_i \Phi(\vec{p})=p_i\Phi(\vec{p}),\end{equation}
\begin{equation}\label{CO}
x_i \Phi(\vec{p})=i(1+\beta p^2)\frac{\partial }{\partial p_i}\Phi(\vec{p}),
\end{equation}
where in our model we have, as before $x_i(i=1,2)=u,v$,
$p_i(i=1,2)=p_u, p_v$ and $p^2=\frac{1}{2}(p_u^2-p_v^2)$. Thus,
one can now define the proper physical states with maximal
localization and use them to define a ``quasi-position
wavefunction". In \cite{7} it is shown that the transition between
the momentum space representation of the wavefunction and its
quasi-position counterpart is given by a generalized Fourier
transformation as
\begin{equation}\label{CP}
\Psi(x)=\sqrt{\frac{2\sqrt{\beta}}{\pi}} \int_{-\infty}^{+\infty}\frac{dp}{(1+\beta p^2)^{3/2}}\exp
\left(\frac{ix \tan^{-1}(\sqrt{\beta}p)}{\sqrt{\beta}}\right)\Phi(p).
\end{equation}
Now, using representations (\ref{CN}) and (\ref{CO}), we get the
following form for the WD equation in momentum space
\begin{equation}\label{CQ}
\left[\frac{\partial^2}{\partial p_v^2}-\frac{\partial^2}{\partial p_u^2}-\frac{\beta}{1+\beta p^2}
\left(p_u \frac{\partial}{\partial p_u}+p_v \frac{\partial}{\partial p_v}\right)+\frac{p^2}{2\omega^2(1+\beta p^2)^2}\right]\Phi(p_u,p_v)=0,
\end{equation}
where in contrast to the minisuperspace coordinate representation of
the WD equation, it is a second order differential equation. This
equation, up to first order in $\beta$ and also neglecting $p^4$ in
the last term takes the form \footnote{Our classical analysis (see
equations (\ref{T})-(\ref{W}), (\ref{AN})-(\ref{AP}) and figure
\ref{fig1}), shows that the dynamical variables $u$ and $v$ have the
same order of magnitude. Therefore the Hamiltonian constraint ${\cal
H}=0$ implies that $p_u$ and $p_v$ should also have the same order
of magnitude. In this sense, neglecting $p^4$ in equation (\ref{CQ})
is quite reasonable.}
\begin{equation}\label{CR}
\left[\frac{\partial^2}{\partial p_v^2}-\frac{\partial^2}{\partial p_u^2}-\beta \left(p_u \frac{\partial}{\partial p_u}+p_v \frac{\partial}{\partial p_v}\right)+\frac{1}{4\omega^2}(p_u^2-p_v^2)\right]\Phi(p_u,p_v)=0,
\end{equation}
which is a separable equation and its solutions may be written as
$\Phi(p_u,p_v)=\Pi(p_u)\Upsilon(p_v)$, leading to
\begin{equation}\label{CS}
\left[\frac{d^2}{dp_u^2}+\beta p_u \frac{d}{dp_u}-\left(\frac{1}{4\omega^2}p_u^2+E\right)\right]\Pi(p_u)=0,
\end{equation}
\begin{equation}\label{CT}
\left[\frac{d^2}{dp_v^2}-\beta p_v \frac{d}{dp_v}-\left(\frac{1}{4\omega^2}p_v^2+E\right)\right]\Upsilon(p_v)=0,
\end{equation}
where $E$ is a separation constant. The above equations have exact
solutions in terms of confluent hypergeometric functions as
\begin{equation}\label{CU}
\Pi_E(p_u)=p_u
\exp\left[-\left(\frac{\sqrt{1+\beta^2\omega^2}}{4\omega}+\frac{\beta}{4}\right)p_u^2\right]U\left(\frac{3}{4}+\frac{\omega(2E+\beta)}{4\sqrt{1+\beta^2\omega^2}},\frac{3}{2};
\frac{\sqrt{1+\beta^2\omega^2}}{2\omega}p_u^2\right),\end{equation}
\begin{equation}\label{CV}
\Upsilon_E(p_v)=p_v
\exp\left[-\left(\frac{\sqrt{1+\beta^2\omega^2}}{4\omega}-\frac{\beta}{4}\right)p_v^2\right]U\left(\frac{3}{4}+\frac{\omega(2E-\beta)}{4\sqrt{1+\beta^2\omega^2}}
,\frac{3}{2}; \frac{\sqrt{1+\beta^2\omega^2}}{2\omega}p_v^2\right).
\end{equation}
Therefore, up to first order in $\beta$ the eigenfunctions of
equation (\ref{CR}) can be written as
\begin{equation}\label{CW}
\Phi_E(p_u,p_v)=p_u p_v e^{-(\frac{1}{4\omega}+\frac{\beta}{4})p_u^2}e^{-(\frac{1}{4\omega}-\frac{\beta}{4})p_v^2}U\left(\frac{3}{4}+\frac{\omega (2E+
\beta)}{4},\frac{3}{2};\frac{p_u^2}{2\omega}\right)U\left(\frac{3}{4}+\frac{\omega (2E-
\beta)}{4},\frac{3}{2};\frac{p_v^2}{2\omega}\right).
\end{equation}
Now the eigenfunctions in the $u-v$ representation can be obtained
from (\ref{CP})
\begin{equation}\label{CX}
\Psi_E(u,v)=\frac{2\sqrt{\beta}}{\pi}\int_{-\infty}^{+\infty}\int_{-\infty}^{+\infty}\frac{dp_u
dp_v}{(1+\beta p^2)^3} \exp
\left(\frac{iu\tan^{-1}(\sqrt{\beta}p_u)}{\sqrt{\beta}}\right)\exp
\left(\frac{iv\tan^{-1}(\sqrt{\beta}p_v)}{\sqrt{\beta}}\right)
\Phi_E(p_u,p_v).
\end{equation}
It is seen that this expression is too complicated for extracting an
analytical closed form for the eigenfunctions in terms of the
minisuperspace variables $u$ and $v$. Moreover, the wavefunction is
a superposition of these eigenfunctions as
\begin{equation}\label{CY}
\Psi(u,v)=\sum_{E}C(E)\Psi_E(u,v).
\end{equation}
On the other hand to construct such a superposition we need to
know whether the separating constant $E$ has a quantized or a
continuum spectrum. Here, we do not intend to deal with such
questions since they are out of the scope of our study. However,
in figure \ref{fig5} we have plotted the square of the
wavefunction where we have taken a discrete superposition of four
terms in (\ref{CY}), all to first order in $\beta$. It should be
noted that we have relaxed the assumption of small values for $u$
and $v$ in this analysis and figure \ref{fig5} shows the square of
the wavefunction for a wide range of these variables. Although
this figure does not completely coincide with figure \ref{fig4},
as is clear,  the wavefunction has its dominant amplitude in
regions represented by the small values of $u$ and $v$, in
agreement with our previous analysis of the GUP quantum model
based on equation (\ref{CI}).
\section{Conclusions}
In this paper we have studied the effects of deformations
(noncommutativity and GUP) in phase space on the cosmic evolution
of a two dimensional minisuperspace model. Our starting point was
the lowest order gravi-dilaton effective action in the string
frame where the Einstein-Hilbert action with a positive
cosmological constant is augmented with an scalar field, the
dilaton. By considering a flat FRW metric for the space time
geometry in this action, we obtained the corresponding effective
Hamiltonian in the minisuperspace constructed by the scale factor
$a$ and dilaton field $\phi$. We saw that this minisuperspace has
a curved metric and thus does not have the desired form for
introducing noncommutativity and GUP between its coordinates. For
this reason, we introduced a new set of variables $(u,v)$, in
terms of which the minisuper metric took a Minkowskian form and
the Hamiltonian of the model described a simple isotropic
oscillator-ghost-oscillator system in which the cosmological
constant plays the role of frequency. Although this is a simple
toy model, these variables are suitable candidates for a
phenomenological study of noncommutativity and GUP in the
corresponding phase space. Another feature of the model in the
$(u,v)$ coordinates is that the duality symmetry of the string
dilaton action exhibits itself as a special linear combination of
$u$ and $v$. In the case of (Poisson) commutative phase space we
saw that both dynamical variables have oscillatory behavior with
the same amplitude. Depending on the initial conditions, these
oscillations may occur in the same or opposite directions. As for
the quantum version of this commutative model, we obtained exact
solutions of the WD equation. The wavefunction of the
corresponding universe consists of two branches where each may be
interpreted as part of the classical trajectory. We saw that since
the peaks of the wavefunction follow the classical trajectory,
there seems to be good correlations between the corresponding
classical and quantum cosmology.

We also studied the noncommutative cosmology where the commutator
between the minisuperspace variables is deformed through a
noncommutative parameter $\theta$. In the classical noncommutative
model this parameter plays the role of a coupling constant between
equations of motion for dynamical variables $u$ and $v$. We solved
the equations of motion exactly and showed that the variables $u$
and $v$ oscillate in the same or opposite directions with an
exponentially damping or increasing amplitude respectively. In the
noncommutative quantum cosmology, we found that the corresponding
Hamiltonian is constructed out of two operators which commute with
each other and thus have the same eigenfunctions. Although the
ensuing WD equation in this case is not separable in the $(u,v)$
coordinates we introduced a new set of variables $(\rho,\varphi)$
in terms of which the WD equation was amenable to exact solutions
in terms of confluent hypergeometric functions. Finally, when the
phase space variables obey the GUP relations we constructed the
classical equations of motion and seen that they form a system of
nonlinear differential equations which cannot be solved
analytically. We showed that the oscillatory behavior of $u$ and
$v$ in the same or opposite directions is again repeated but in
this case the oscillations are not harmonic. We also found that
the GUP parameter $\beta$ causes a larger period in comparison to
the ordinary model where $\beta=0$. This can be interpreted as the
importance of quantum gravitational effects not only at early
times but also at late times of the cosmic evolution. The
resulting quantum cosmology and the corresponding WD equation in
the GUP framework were also studied and approximate analytical
expressions for the wavefunctions of the universe were presented
in the limit of small $u$ and $v$ variables. These solutions show
only one possible state with no classical correlation at early
times from which our universe can emerge. We saw that such
behavior also occurs in the noncommutative quantum model, showing
that from the point of view adopted here, noncommutativity and GUP
may be considered as similar concepts.


\begin{thebibliography}{99}
\bibitem{1}S.W. Hawking and G.F.R. Ellis, {\it The Large Scale Structure of Space-Time}, Cambridge University Press, Cambridge, 1973\\L.D. Landau and
E.M. Lifshitz, {\it The Classical Theory of Fields}, Pergamon Press, Oxford, 1975\\M.P. Ryan and L.C. Shepley, {\it Homogeneous Relativistic Cosmologies}, Princeton University Press, Princeton, 1975\\J.N. Islam, {\it An Introduction to Mathematical Cosmology}, Cambridge University Press, Cambridge, 2001
\bibitem{2}B.S. DeWitt, {\it Phys. Rev.} {\bf 160} (1967) 1113\\C.W. Misner, {\it Phys. Rev.} {\bf 186} (1969) 1319\\J.B. Hartle and S.W. Hawking,
{\it Phys. Rev.} D {\bf 28} (1983) 2960\\S.W. Hawking, {\it Nucl. Phys.} B {\bf 239} (1984) 257\\S.W. Hawking, {\it Phys. Rev.} D {\bf 37} (1988) 904\\S.W. Hawking and D. Page, {\it Nucl. Phys.} B {\bf 264} (1986) 185
\bibitem{3} D. J. Gross and P. F. Mende, {\it Nucl. Phys.} B {\bf
303}, (1988) 407\\ D. Amati, M. Ciafaloni and G. Veneziano, {\it Phys.
Lett.} B {\bf 216} (1989) 41\\ M. Kato, {\it Phys. Lett.} B {\bf 245} (1990) 43\\ S. de Haro, {\it J. High Energy Phys.} {\bf JHEP 9810}
(1998) 023\\ L. G. Garay, {\it Int. J. Mod. Phys.} A {\bf 10}
(1995) 145 (arXiv: gr-qc/9403008)\\ K. Konishi, G. Paffuti and P. Provero, {\it Phys.
Lett.} B {\bf 234} (1990) 276
\bibitem{4}A. Ashtekar and J. Lewandowski, {\it Class. Quantum Grav.} {\bf 21} (2004) R53\\C. Rovelli, {\it Quantum Gravity},
Cambridge University Press, Cambridge, 2004
\bibitem{5} S. Capozziello, G. Lambiase and G. Scarpetta, {\it
Int. J. Theor. Phys.} {\bf 39} (2000) 15
\bibitem{6} F. Scardigli, {\it Phys. Lett.} B {\bf 452} (1999) 39
(arXiv: hep-th/9904025)\\K. Nozari and B. Fazlpour, {\it  Mod. Phys. Lett.} A {\bf 38} (2007) 2917 (arXiv: hep-th/0605109)\\K. Nozari and B. Fazlpour, {\it Acta Phys. Pol.} B {\bf 39} (2008) 1363 (arXiv: gr-qc/0608077)\\K. Nozari and S.H. Mehdipour, {\it  Failure of Standard Thermodynamics in Planck Scale Black Hole System} (arXiv: hep-th/0610076)
\bibitem{7} A. Kempf, G. Mangano and R. B. Mann, {\it Phys. Rev.}
D {\bf 52} (1995) 1108 (arXiv: hep-th/9412167)
\bibitem{8} A. Kempf and G. Mangano, {\it Phys. Rev.} D {\bf 55}
(1997) 7909 (arXiv: hep-th/9612084)
\bibitem{9} M. Maggiore, {\it Phys. Lett.} B {\bf 319} (1993)
83\\A. Kempf, {\it J. Math. Phys.} {\bf 35} (1994) 4483\\A. Kempf and J. C. Niemeyer, {\it Phys. Rev.} D {\bf 64}
(2001) 103501\\L. N. Chang, D. Minic, N. Okamura and T. Takeuchi,
{\it Phys. Rev.} D {\bf 65} (2002) 125027
\bibitem{10}H. Snyder, {\it Phys. Rev.} {\bf 71} (1947) 38
\bibitem{11}A. Connes, {\it Noncommutative Geometry}, Academic, New
York, 1994\\A. Connes {\it J. Math. Phys.} (N.Y.) {\bf 41} (2000) 3832\\ J.C. Varilly, {\it  An Introduction to Noncommutative Geometry} (arXiv: physics/9709045)\\
M.R. Douglas and N.A. Nekrasov, {\it Rev. Mod. Phys.} {\bf 73} (2001) 977
\bibitem{12}S.F. Hassan and M.S. Sloth, {\it Nucl. Phys.} B {\bf 674} 434 (arXiv: hep-th/0204110)
\bibitem{13}B. Vakili, N. Khosravi and H.R. Sepangi, {\it Class. Quantum Grav.} {\bf 24} (2007) 931 (arXiv: gr-qc/0701075)\\
N. Khosravi, H.R. Sepangi and M.M. Sheikh-Jabbari, {\it Phys. Lett.} B {\bf 647} (2007) 219 (arXiv: hep-th/0611236)\\ N. Khosravi, S. Jalalzadeh and H.R. Sepangi, {\it Gen. Rel. Grav.} {\bf 39} (2007) 899 (arXiv: gr-qc/0702067)
\bibitem{14}W. Guzman, M. Sabido and J. Socorro, {\it On Noncommutative Minisuperspace and the Friedmann equations} (arXiv: 0812.4251 [gr-qc])\\
H. Garcia-Compean, O. Obregon and C. Ramirez, {\it Phys. Rev. Lett.} {\bf 88} (2002) 161301 (arXiv: hep-th/0107250)\\
Y.-F. Cai and Y.-S. Piao, {\it Phys. Lett.} B {\bf 657} (2007) 1 (arXiv: gr-qc/0701114)\\C. Bastos, O. Bertolami, N.C. Dias and J.N. Prata, {\it Phys. Rev.}D {\bf 78} (2008) 023516 (arXiv: 0712.4122 [gr-qc])
\bibitem{15}M.V. Battisti and
G. Montani, {\it Phys. Lett.} B {\bf 656} (2007) 96 (arXiv: gr-qc/0703025)\\M.V. Battisti and G. Montani, {\it Phys. Rev.} D {\bf 77} (2008) 023518 (arXiv: 0707.2726 [gr-qc])\\
A. Bina, K. Atazadeh and S. Jalalzadeh, {\it Int. J. Thoer. Phys.} {\bf 47} (2008) 1354 (arXiv: 0709.3623 [gr-qc])\\M.V. Battisti and G. Montani, {\it AIP Conf. Proc.} {\bf 966} (2008)
219 (arXiv: 0709.4610 [gr-qc])\\M.V. Battisti and G. Montani, {\it Int.
J. Mod. Phys.} A {\bf 23} (2008) 1257 (arXiv: 0802.0688[gr-qc])
\bibitem{16}B. Vakili and H.R. Sepangi, {\it Phys. Lett.} B {\bf 651} (2007) 79 (arXiv: 0706.0273 [gr-qc])
\bibitem{17} B. Vakili, {\it Phys. Rev.} D {\bf 77} (2008) 044023 (arXiv: 0801.2438 [gr-qc])
\bibitem{18}G. Veneziano, {\it Phys. Lett.} B {\bf 265} (1991)
287\\M. Gasperini and G. Veneziano, {\it Astropart. Phys.} {\bf 1} (1993) 317\\ V. Bozza, M. Gasperini, M. Giovannini and G. Veneziano, {\it  Phys. Lett.} B {\bf 543} (2002) 14\\ V. Bozza, M. Gasperini, M. Giovannini and G. Veneziano, {\it Phys. Rev.} D {\bf 67} (2003) 063514

\bibitem{19}M. Gasperini, {\it Dilaton Cosmology and Phenomenology}, (arXiv: hep-th/0702166)
\bibitem{20}S. Capozziello and R. de Ritis, {\it Int. J. Mod. Phys.}
D {\bf 2} (1993) 373\\E.J. Copeland, A. Lahiri and D. Wands, {\it
Phys. Rev.} D {\bf 50} (1994) 4868\\S. Capozziello, G. Lambiase and
R. Capaldo, {\it Int. J. Mod. Phys.} D {\bf 8} (1999) 213 (arXiv: gr-qc/9805046)
\bibitem{21}M. Gasperini, J. Maharana and G. Veneziano, {\it Nucl. Phys.} B {\bf 472} (1996) 349\\
U.H. Danielsson, {\it Class. Quantum Grav.} {\bf 22} (2005) S1-S40 (arXiv: hep-th/0409274)
\bibitem{22} A. Vilenkin, {\it Phys. Rev.} D {\bf 37} (1988) 888
\bibitem{Mo}J.E. Moyal, {\it Proc. Cambridge Phil. Soc.} {\bf 45} (1949) 99\\
A.C. Hirshfeld and P. Henselder, {\it Am. J. Phys.} {\bf 70} (2002) 5 (arXiv: quant-ph/0208163)
\bibitem{Za}C. Zachos, {\it Int. J. Mod. Phys.} A {\bf 17} (2002) 297 (arXix: hep-th/0110114)
\bibitem{Car}J.M. Carmona, J.L. Cortes, J. Gamboa and F. Mendez, {\it J. High Energy Phys.} {\bf JHEP 0303} (2003)
058 (arXiv: hep-th/0301248)\\J.M. Carmona, J.L. Cortes, J. Gamboa and F. Mendez, {\it Phys. Lett.} B {\bf 565} (2003) 222 (arXiv:
hep-th/0207158)
\bibitem{Rom}J.M. Romero, J.D. Vergara and J.A. Santiago, {\it Phys. Rev.} D {\bf 75} (2007) 065008, (arXiv: hep-th/0702113)\\
N. Khosravi and H.R.Sepangi, {\it  Phys. Lett.} A {\bf 372} (2008) 3356 (arXiv: 0802.0767 [gr-qc])\\N. Khosravi and H. R. Sepangi, {\it J. Cosmol. Astropart. Phys.} {\bf JCAP 0804} (2008) 011 (arXiv: 0803.1714 [gr-qc]
\bibitem{23}S. Benczik, et al., {\it Phys. Rev.} D
{\bf 66} (2002) 026003 (arXiv: hep-th/0204049)\\S. Benczik, et al.,
{\it Classical Implications of the Minimal Length Uncertainty
Relation} (arXiv: hep-th/0209119)
\bibitem{24} K. Nozari and S. Akhshabi, {\it  On the Stability of Planetary Circular Orbits in Noncommutative
Spaces} (arXiv: gr-qc/0608076)\\ K. Nozari and S. Akhshabi, {\it
Europhys. Lett.} {\bf 80} (2007) 20002 (arXiv: 0708.3714[gr-qc])
\bibitem{25}A. Vilenkin, {\it Phys. Rev.} D {\bf 37} (1988) 888
\bibitem{26}M. Abramowitz and I.A. Stegun, {\it Handbook of Mathematical Functions} (1972) (New York: Dover)
\end{thebibliography}
\end{document}